\documentclass[11pt]{article}
\pdfoutput=1
\usepackage{graphicx,amssymb,amsmath}
\graphicspath{{figs/}}

\usepackage[colorlinks=true,urlcolor=blue,
citecolor=blue,linkcolor=blue,pdfa=true]{hyperref}

\begin{document}

\begin{center}\large\bfseries
	Search for extreme energy cosmic ray candidates in the TUS orbital
	experiment data
\end{center}

\begin{center}
S.V.~Biktemerova$^{b}$,
A.A.~Botvinko$^{c}$,
N.P.~Chirskaya$^{a}$,
V.E.~Eremeev$^{a}$,
G.K.~Garipov$^{a}$,
V.M.~Grebenyuk$^{b,d}$,
A.A.~Grinyuk$^{a}$,
S.~Jeong$^{f}$,
N.N.~Kalmykov$^{a}$,
M.A.~Kaznacheeva$^{a}$,
B.A.~Khrenov$^{a}$,
M.~Kim$^{f}$,
P.A.~Klimov$^{a}$,
M.V.~Lavrova$^{a}$,
J.~Lee$^{f}$,
O.~Martinez$^{g}$,
M.I.~Panasyuk$^{a}$,
I.H.~Park$^{f}$,
V.L.~Petrov$^{a}$,
E.~Ponce$^{g}$,
A.E.~Puchkov$^{c}$,
H.~Salazar$^{g}$,
O.A.~Saprykin$^{c}$,
A.N.~Senkovsky$^{c}$,
S.A.~Sharakin$^{a}$,
A.V.~Shirokov$^{a}$,
A.V.~Tkachenko$^{b}$,
L.G.~Tkachev$^{b,d}$,
I.V.~Yashin$^{a}$,
M.Yu.~Zotov$^a$\\
(The Lomonosov--UHECR/TLE collaboration)
\end{center}

\bigskip
{\narrower
\par\noindent{$^a$M.V. Lomonosov Moscow State University, GSP-1, Leninskie
	Gory, Moscow, 119991,\\ Russia}
\par\noindent{$^b$Joint Institute for Nuclear Research, Joliot-Curie, 6,
	Dubna, Moscow region, Russia, 141980}
\par\noindent{$^c$Space Regatta Consortium, ul. Lenina, 4a,
	141070 Korolev, Moscow region, Russia}
\par\noindent{$^d$Dubna State University, University str., 19, Bld.1,
	Dubna, Moscow region, Russia}
\par\noindent{$^e$Department of Physics and ISTS, Sungkyunkwan
	University,	Seobu-ro 2066, Suwon,\\ 440-746 Korea}
\par\noindent{$^f$Benem\'{e}rita Universidad Aut\'{o}noma de Puebla,
	4 sur 104 Centro Hist\'orico C.P. 72000, Puebla, Mexico}

}

\begin{abstract}

	TUS (Track Ultraviolet Setup) is the first space experiment aimed to
	check the possibility of registering extreme energy cosmic rays
	(EECRs) at $E>50$~EeV by measuring the fluorescence signal of
	extensive air showers in the atmosphere.  The detector operates as a
	part of the scientific payload of the Lomonosov satellite for more
	than a year.
	We describe an algorithm of searching for EECR events in the TUS data
	and briefly discuss a number of candidates selected by formal
	criteria.

\end{abstract}

\section{Introduction}

Measurements of the cosmic ray (CR) energy spectrum, their nuclear
composition and arrival directions at extreme energies
$E\gtrsim5\times10^{19}$~eV are an important part of
modern astrophysics and particle physics~\cite{Dawson_etal-2017}.

The first CR particles with energy $E\gtrsim50$~EeV were
detected~\cite{1961PhRvL...6..485L} and a cut-off of the energy spectrum
was predicted~\cite{Greisen-1966,ZK-1966} more than 50 years ago but
results in the EECR study do not give clear answers to the most
important question of the Greisen--Zatsepin--Kuz'min
cut-off~\cite{Berez2014}. The nature and origin of EECRs are still not
understood. To a great extent, the problem relates to a very low flux of
EECRs: the two largest ground-based arrays---The Pierre Auger
Observatory and the Telescope Array---registered less than two dozen
events with energies $E>100$~EeV in 10 and 5 years of operation
respectively~\cite{PierreAuger:2014yba,Abbasi:2014lda}.

A primary goal of the TUS project, first announced in
2001~\cite{2001AIPC..566...57K}, is to expand the EECR experimental
studies to space as was suggested by Benson and Linsley in early
1980's~\cite{1980BAAS...12Q.818B,Benson-Linsley-1981}. The main idea is
that fluorescent and Cherenkov radiation of an extensive air shower
(EAS) generated by an EECR in the nocturnal atmosphere of the Earth can
be detected from a satellite similar to the way it is observed from the
ground but with a much larger exposure, thus considerably increasing the
statistics of registered events.

Following the idea, a number of orbital detectors with a large signal
collecting area and high time-lateral resolution are being
elaborated~\cite{ECRS}.  The TUS detector is a pathfinder to the
large-scale missions like KLYPVE~\cite{klypve-2015} or
JEM-EUSO~\cite{2014AdSpR..53.1499E}.  Accurate estimations of the energy
and arrival directions of primary EECR particles are left to the future
orbital detectors.

Skobeltsyn Institute of Nuclear  Physics of Lomonosov Moscow State
University (SINP MSU), Joint Institute of Nuclear Research (JINR) and
Space Regatta Consortium together with several Korean and Mexican
Universities have collaborated in the TUS detector preparation. Results
of the detector simulations, development and preflight tests are
published elsewhere~\cite{Khrenov2002,Khrenov2004,Abrashkin2006,Khrenov2013EPJ,Adams2015}.

The Lomonosov satellite was launched into orbit from the newly built
Vostochny Cosmodrome (Russia) on April~28, 2016.
The satellite has a sun-synchronous orbit with an inclination
of $97^\circ\!\!.3$, a period of $\approx94$~min, and a height of about
470--500~km.

\section{The TUS detector data acquisition system and on-line trigger}

\begin{figure}[!ht]
	\centerline{\includegraphics[width=30pc,angle=0]{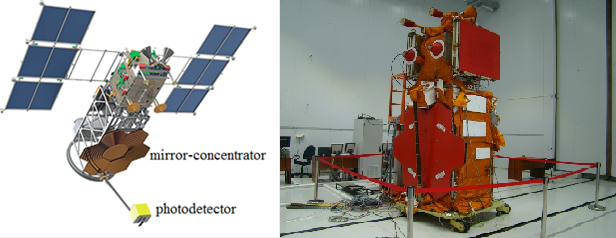}}
	\caption{Left: artist's view of the TUS detector on-board the
	Lomonosov satellite.  Right: the TUS detector assembled on the
	Lomonosov frame before the launch.}
	\label{fig1}
\end{figure}

The TUS detector on board Lomonosov satellite is presented in
Fig.~\ref{fig1}. It consists of the two main parts: a modular Fresnel
mirror-concentrator and a photo-receiver matrix consisting of 256
channels (16 photodetector modules (PDMs) of 16 channels each) in the
focal plane of the mirror.  A sensor of each channel (pixel) is a
Hamamatsu-type R1463 photomultiplier tube (PMT) with a 13~mm diameter
multi-alcali cathode covered by a ultraviolet (UV) glass filter and a
mirror light guide with square entrance of 15~mm size.  Quantum
efficiency of PMTs in the UV band is $\sim20$\%.  The Fresnel mirror has
an area of 2.0~m$^2$ and a focal distance of 1.5~m.  One pixel has an
FOV of 10 mrad, which corresponds to a spatial spot of about
$5~\text{km}\times5~\text{km}$ at the sea level for a 500~km orbit
height.  Thus, the total FOV equals ($\pm4.5^\circ$), and the full area
observed by TUS at any moment is approximately
$80~\text{km}\times80~\text{km}$.  A detailed description of the TUS
detector can be found in~\cite{SSR}.

The general design of TUS is determined by its main task of registering
fluorescence and scattered (reflected) Cherenkov radiation of EASs in the UV
band with a time resolution of 0.8~$\mu$s in full temporal interval of
256 time steps.  It can also be employed for measuring transient events
in the atmosphere with longer duration in modes with longer time samples
implemented in the detector DAQ system.  This design makes TUS a
multi-purpose
detector~\cite{TUS-TEPA-2016,uhecr2016,2017arXiv170407704K}.

A crucial part of any detector is its trigger system. TUS has a
two-level trigger~\cite{2013JPhCS.409a2105G,TUS-sim-ApP-2017}.
The first-level trigger is a threshold trigger: the PDM board
calculates a moving sum (MS) of PMT signals during 16 time
steps in each channel and looks for an MS value above a threshold level.
The threshold level is calculated as $\langle MS\rangle+dQ$, where
$\langle MS\rangle$ is the mean value of MS obtained during previous
100~ms for each channel, and~$dQ$ is a constant unique for all channels.
It can be changed by commands from the mission control center (MCC).
The PDM board transmits a trigger word to the central processor board
(CPB), which forms a map of active pixels with moving sums above the
threshold. 
The second-level trigger is a pixel-mapping trigger implemented in the
CPB. It acts as a contiguity trigger. This procedure selects cases of
sequential triggering of spatially contiguous active pixels that are
also adjacent in time, allowing for the selection of events with a
special spatial-temporal pattern. An additional parameter important for
this trigger is the so-called adjacency length ($L$), i.e., the number of
neighboring channels sequentially activated by a signal from a
given event.
Data of all 256 channels for 256 time steps is transmitted from the CPB
memory to the on-board computer in case conditions of the trigger system
are satisfied.

For the maximal EAS trigger rate of about 1~min$^{-1}$ (confined by the
design of the on-board computer) and the event size of
128~KiB, an amount of data collected in one satellite revolution does not
exceed 12~MiB. This information is transferred to the MCC twice per
day at sessions of the best connection between the satellite and the MCC.

\section{Expected EAS signal: simulation and criteria}
\label{sec:signal}

Various aspects of TUS' operation in space were simulated before the
launch~\cite{2011PPNL....8..789B,2013JPhCS.409a2105G,sim-icrc2015,TUS-sim-ApP-2017}.
The simulations employed ESAF (EUSO Simulation and Analysis
Framework)~\cite{ESAF} and the TUSSIM program developed at JINR.  The
QGS model parametrization of Ilina et~al.~\cite{Ilina} was selected as
the closest to experimental data at primary energies in the range of
10--100~EeV.  It was found in particular that duration of a horizontal EAS
in the FOV of one pixel is $\sim20~\mu$s.  For vertical EASs, it is
approximately two times larger.  The total duration of an EAS signal
depends on the zenith angle and varies approximately from 30~$\mu$s to 
100~$\mu$s.  A measure of the energy of a primary particle is a number
of photons in the EAS maximum, which is related to the electron
(positron) number~$N_\mathrm{e}$ as $N_\mathrm{ph} = Y N_\mathrm{e}$,
where~$Y\sim4.5$~photons/m is a fluorescent photon yield.
The detected amplitude of fluorescence at the
cascade maximum is expected to be of the order of 20 photons/$\mu$s for
a horizontal EAS generated by a 100~EeV proton.

\begin{figure}[!ht]
	\centering
	\includegraphics[width=.6\textwidth]{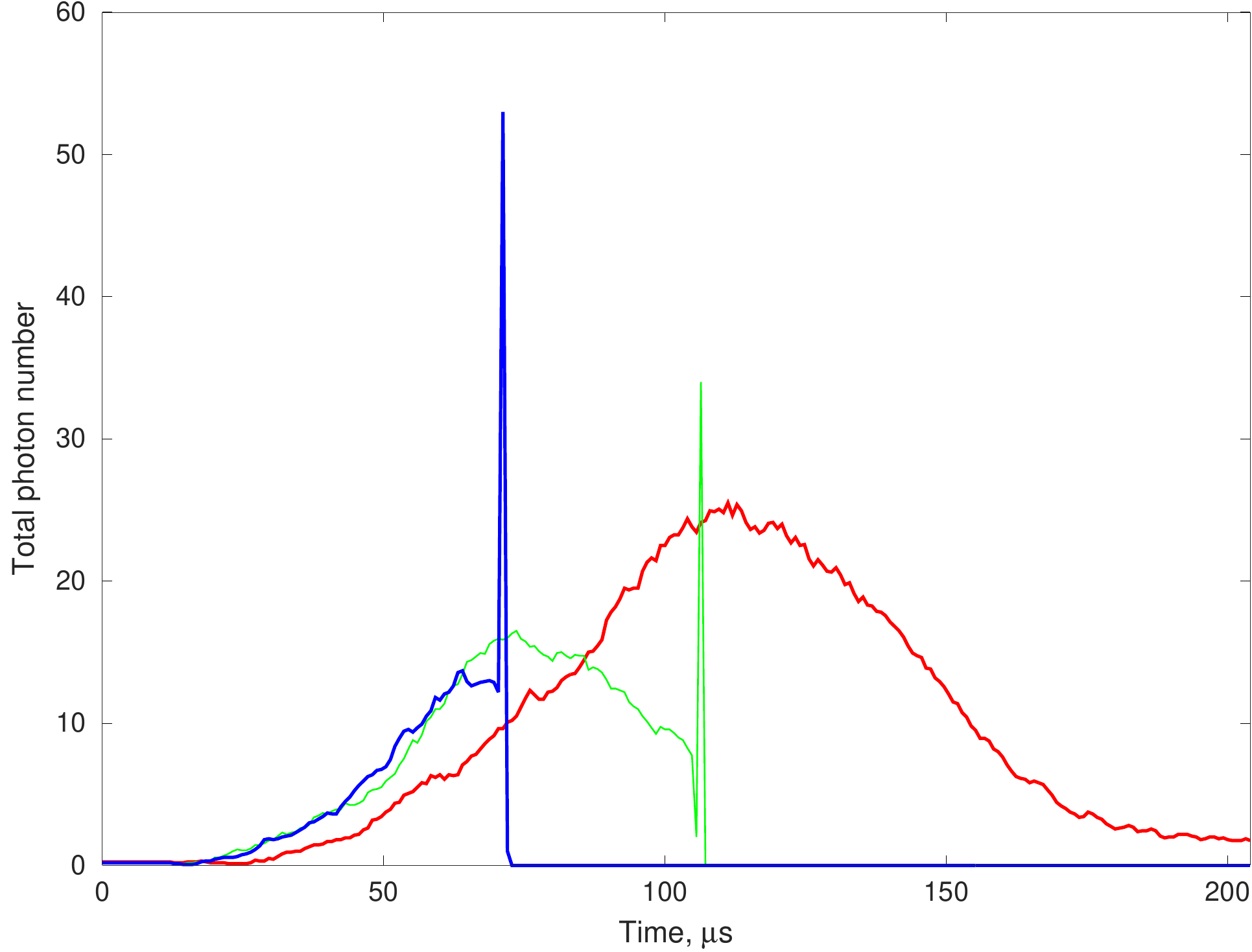}

	\caption{Examples of light curves of an EAS generated by a 100~EeV
	primary proton arriving at three different zenith angles:
	$\theta=15^\circ$ (blue), $\theta=45^\circ$ (green) and
	$\theta=75^\circ$ (red). Zero background radiation is assumed.
	}

	\label{EAS3}
\end{figure}

It should be noticed that maximum of the vertical cascade curve at such
high energies is close to the sea level, so that a considerable part of
the cascade curve escapes observation in the atmosphere.  On the other
hand, it is followed by a strong Cherenkov light signal reflected by the
surface (up to 80\% by snow). A back-scattered Cherenkov ``point'' with
an amplitude of the signal in one time sample of the detector much
higher than the background is a characteristic signal of an EAS.
Typical light curves (total number of UV photons per
microsecond on the detector focal surface as a function of time) for
three values of EAS zenith angle are shown in Fig.~\ref{EAS3}
for a primary proton with the energy 100~EeV approaching the FOV of TUS
along one of the diagonals. The curves were obtained with ESAF and
optics simulation software based on a realistic
3D TUS mirror model, developed at JINR.

The main features of the light curves for (quasi-)vertical and (quasi-)horizontal
EASs are obviously different.
A sharp break of the fluorescent signal and a Cherenkov peak at the
meeting point of an EAS front with the scattering surface (ocean,
continent) are expected for vertical events.
Duration of the signal in one pixel is larger than for
inclined events and the amplitude of the fluorescent signal is
lower.

An image of an EAS on the focal surface produces a moving track.
The speed of the movement depends on the zenith angle~$\theta$
as $\chi = c\tan(\theta/2)/R$, where~$\chi$ is an angular speed of the source,
$R$~is the distance from the source to the detector (which weakly
depends on~$\theta$), and~$c$ is the speed of light in the atmosphere.
This leads to the difference in the signal duration in different channels.
A comparison of parameters of EASs with various zenith angles is
presented in Table~\ref{Table1}. Here, $T_\text{pix}$ is
duration of the signal in one pixel estimated as $\gamma_\text{pix}/\chi$
with $\gamma_\text{pix}=10$~mrad being the FOV of one pixel, and
$T_{1/2}$ is the full duration at half maximum (FDHM) of the
light curve. All calculations were made for a 100~EeV primary proton.

\begin{table}[!ht]
\caption{Typical parameters of EASs with different zenith angles:
	an altitude of maximum of the cascade $H_\text{max}$,
	a distance from TUS to the EAS maximum $R_\text{max}$,
	an angular speed of the signal source $\chi$,
	duration of the signal in one channel $T_\text{pix}$,
	and the FDHM $T_{1/2}$ of the EAS light curve.}

\begin{center}
	\begin{tabular}{|c|c|c|c|c|c|}
		\hline
		$\theta$ & $H_\text{max}$,~km &
		$\chi$,~mrad/$\mu$s & $T_\text{pix},~\mu$s & $T_{1/2},~\mu$s \\
		\hline
		15$^\circ$ &1.5  & 0.08& 123& $<35$\\
		30$^\circ$ &2.5  & 0.17& 61 &38\\
		45$^\circ$ &4.2  & 0.26& 39& 42\\
		60$^\circ$ &6.3  & 0.36& 28& 47\\
		75$^\circ$&11 & 0.48& 21& 69\\
		80$^\circ$&14  & 0.53 & 19& 94\\
		\hline
	\end{tabular}
\end{center}
\label{Table1}
\end{table}

It is clear from the table that
the temporal evolution of the signal strongly differs
for the two limiting intervals of the zenith angle.
For the sake of brevity, we shall call air showers with $\theta < 30^\circ$
as vertical, those with $\theta>60^\circ$ as horizontal, and all the
rest as inclined.

\begin{figure}[!ht]
	\centering
	\includegraphics[width=.6\textwidth]{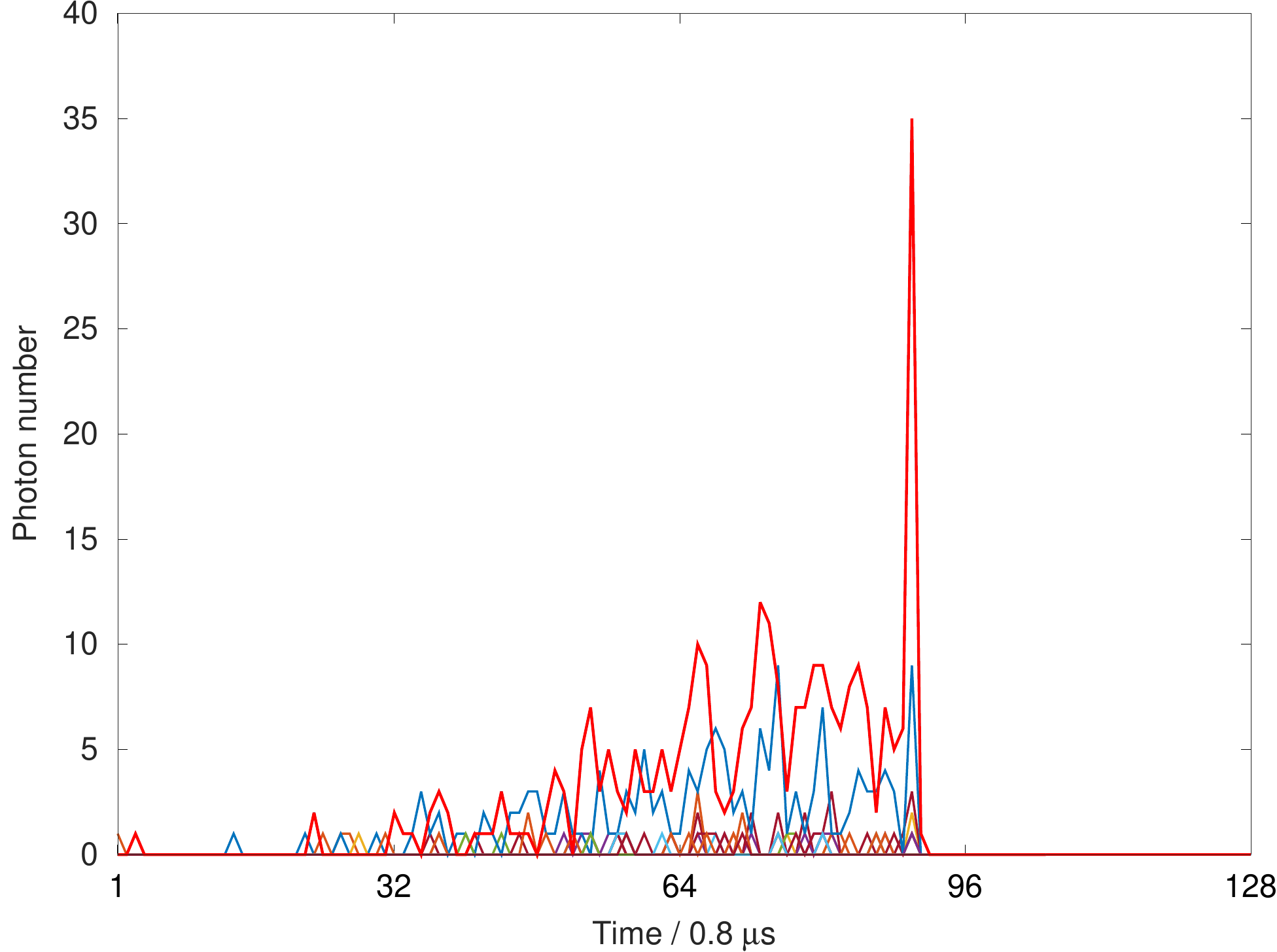}

	\caption{Flux of photons on the TUS focal surface
	for a simulated vertical EAS event from a 100~EeV proton.
	Colors denote different channels.
	The Cherenkov peak at the end coresponds to the albedo of~0.05.
	Zero background is assumed.}

	\label{fig:vert}
\end{figure}

\begin{figure}[!ht]
	\centering
	\includegraphics[width=.6\textwidth]{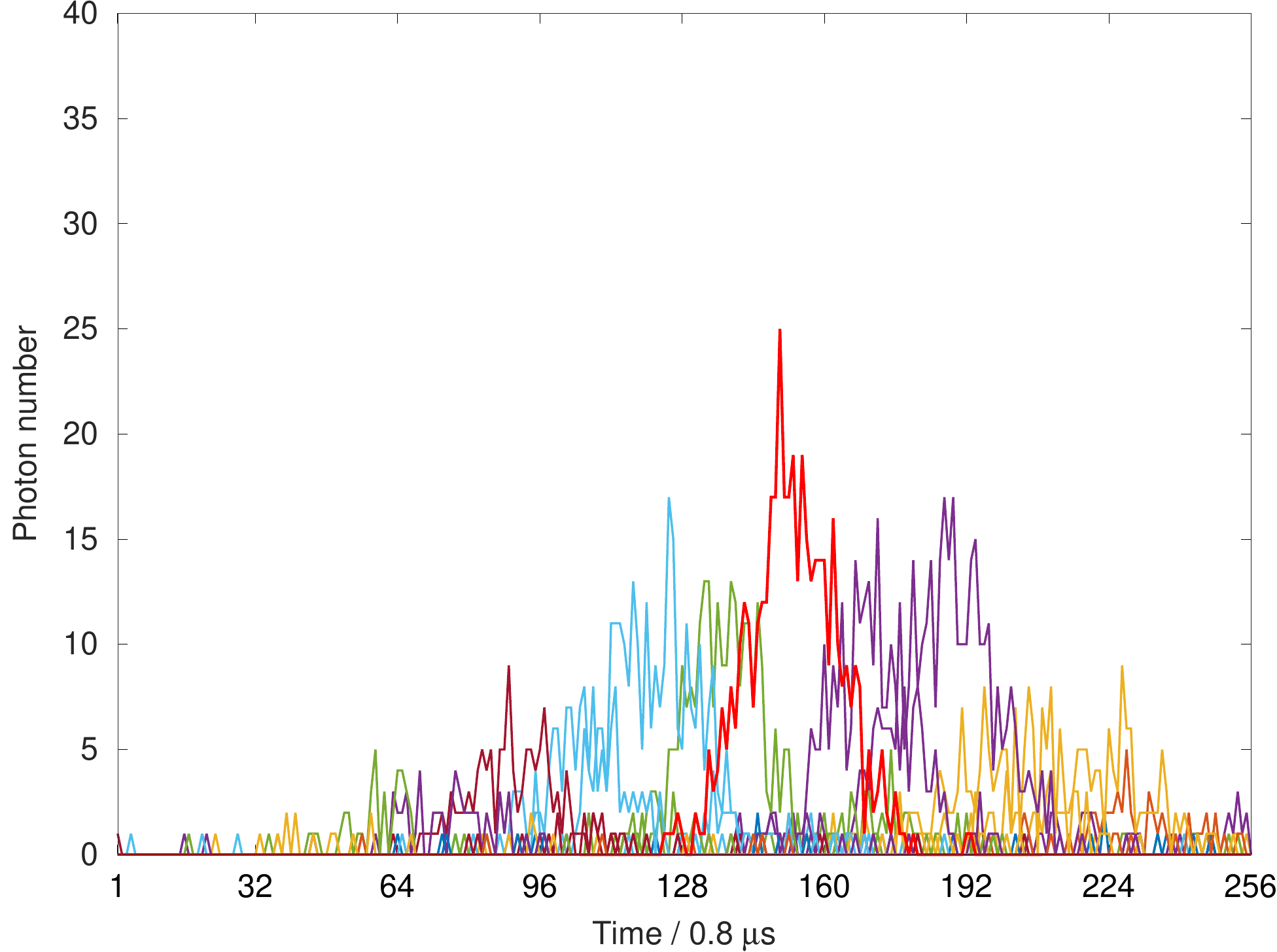}

	\caption{Flux of photons on the TUS focal surface
	for a horizontal EAS from a 100~EeV proton.
	Colors denote different channels.}
	\label{fig:hor}
\end{figure}

Figures~\ref{fig:vert} and~\ref{fig:hor} demonstrate results of
simulations of the flux of photons in the TUS detector channels for a
vertical ($\theta=15^\circ$) and horizontal ($\theta=75^\circ$) air
showers respectively. Zero background radiation is assumed. In both
cases, the EASs are generated by a 100~EeV primary proton arriving along
one of the diagonals of the FOV.
One can see features discussed above for the vertical EAS shown:
the duration of the signal in the brightest pixel is $\approx50~\mu$s
with a very short ($<1~\mu$s) flash of the
Cherenkov light at the end. This vertical EAS signal is expected to be
measured in one or two pixels of an ``ideal'' detector with a small
point spread function (PSF). In the real TUS detector, the PSF spreads
for more than one pixel, and the signal might be registered in a group
of 3--5 adjacent pixels.

A horizontal EAS shown in Fig.~\ref{fig:hor} demonstrates a completely
different signal.  In contrast to the vertical EAS, signals in separate
channels have shorter duration with $\sim20$~photons/$\mu$s in the
brightest pixels.  Information that can be obtained from inclined
showers is richer than from vertical ones due to the signal detection in
several pixels and larger signal-to-noise ratio due to shorter duration
of useful time interval for EAS photons collection.
An analysis of the light curves of different types of EASs allow one to
obtain several criteria for selecting EECR candidates in
the TUS data, see below.

\section{EAS candidates in the TUS data}

TUS began measurements in space on May 19, 2016. The first months of
work were mostly dedicated to testing hardware, software and various
modes of operation. Continuous measurements were started in September
2016 with several gaps for the satellite technical service.
By the end of May 2017, TUS registered around 35,000 events at
nocturnal parts of its orbit in the EAS mode.
The total exposure is approximately $530$~km$^2$~yr~sr.
It is small in comparison with that of the largest
ground-based arrays but sufficient to look for an EECR above
the TUS threshold energy $E\gtrsim70$~EeV~\cite{TUS-sim-ApP-2017}.

In a search for possible EAS candidates, we analyzed approximately 10,000
events registered in the conditions of the minimal background radiation
and considered those that satisfied the following criteria. First, we
selected events such that the moving sum of a waveform in the event
exceeded the background level by at least 96 ADC counts (i.e., $dQ=96$),
and the adjacency length $L\ge6$ (i.e., satisfied conditions of a
``software'' trigger).
The number of such software triggers in an event should be less than~17.

Among events obtained at the first stage, we selected those that satisfied
additional constraints based on the Gaussian approximation of waveforms
of active pixels (``fit conditions''):
\begin{itemize}
	\item the peak of the signal was located within 72\dots230~$\mu$s
		from the beginning of the record;

	\item total duration of the signal in any active channel was within
		27\dots144~$\mu$s;

	\item the coefficient of multiple determination $R^2>0.8$.
	
\end{itemize}

The above criteria were developed basing on simulations of a few
thousand EECRs with different parameters of primary particles
and an analysis of different types of ``noise'' events in the data
set~\cite{2017arXiv170407704K}.
The procedure led us to a list of 13 events presented in
Table~\ref{table2}.

\begin{table}[!ht]
	\caption{EAS candidates and their parameters. Date is given in the
	YYMMDD format.
	T\&F is the number of active pixels that met both the software
	trigger and fit criteria.}

	\begin{center}
	\begin{tabular}{|c|c|c|c|c|c|c|}
		\hline
		\#	&Date&	UTC&	Geolocation&	T\&F\\
		\hline
		1&	160928&	06:30:53&	$38^\circ\!\!.4$N, $104^\circ\!\!.4$W&	12\\
		2&	160930&	05:39:02&	$30^\circ\!\!.9$N, $92^\circ\!\!.9$W&	1	\\
		3&	161003&	05:48:59&	$44^\circ\!\!.1$N, $92^\circ\!\!.7$W&	6	\\
		4&	161003&	05:50:02&	$40^\circ\!\!.1$N, $93^\circ\!\!.9$W&	1	\\
		5&	161023&	07:49:56&	$41^\circ\!\!.3$N, $123^\circ\!\!.8$W&	2	\\
		6&	161024&	10:25:57&	$61^\circ\!\!.4$N, $155^\circ\!\!.8$W&	1 \\	
		7&	161031&	10:25:18&	$61^\circ\!\!.3$N, $155^\circ\!\!.7$W&	4	\\
		8&	161124&	20:20:19&	$13^\circ\!\!.5$S, $40^\circ\!\!.2$E&	1	\\
		9&	161208&	07:20:40&	$74^\circ\!\!.7$N, $95^\circ\!\!.6$W&	1	\\
		10&	170420&	02:18:33&	$12^\circ\!\!.4$S, $50^\circ\!\!.4$W&	2\\
		11&	170421&	17:22:26&	$27^\circ\!\!.7$N, $89^\circ\!\!.0$E&	1\\
		12&	170422&	13:52:43&	$4^\circ\!\!.9$N, $138^\circ\!\!.2$E&	1\\
		13&	170422&	16:51:52&	$39^\circ\!\!.9$N, $98^\circ\!\!.9$E&	2\\
		\hline
	\end{tabular}
	\end{center}
	\label{table2}
\end{table}

The next step of the analysis is an event by event study of
the temporal and spatial dynamics of the signal.
This analyses is aimed to search for
typical signatures of an EAS taking into account the characteristic
temporal parameters, signal amplitudes and image structure discussed
above. The latter one
is mainly determined by the PSF, which is sufficiently larger than a pixel
size and may differ in various parts of the FOV.
The preflight measurements of the PSF
showed that is has a diameter of 27-30~mm (70\% of energy) on the edge
of the FOV, which twice larger than the size of a pixel~\cite{2014NIMPA.763..604G}.
Simple estimations show that
it can increase the  signal duration in 1.5--2 times in comparison with
an ideal optical system. For a more accurate determination of the
temporal parameters of a signal (duration in a separate channel and the
total duration of the light curve), their evaluation was made by means
of the FDHM.

As was shown in Section~\ref{sec:signal}, it is expedient for the
subsequent analysis to formulate two different sets of the second level
criteria, separately for horizontal and vertical events.

\textit{Horizontal EAS criteria:}
\begin{enumerate}

	\item Linearity of the image: alignment of 5 or more active pixels
		in a line (``track'').

	\item The characteristic duration of the signal in one pixel:
		20--50~$\mu$s.

	\item The characteristic total duration of the signal: more than
		50~$\mu$s.

\end{enumerate}

\textit{Vertical EAS criteria:}
\begin{enumerate}

	\item Spot-like image: the signal is confined to 2--4 neighboring channels.

	\item The characteristic duration of the signal in one channel:
		more than 40~$\mu$s.

	\item The characteristic total duration of the signal: less than
		50~$\mu$s.

	\item Presence of the Cherenkov peak at the end of the light curve
		during one time step and not later than in 10--30~$\mu$s after the
		maximum of the EAS fluorescence.

\end{enumerate}

As a result of this event by event analysis based on the second stage
criteria, the preliminary EAS candidates were classified as follows:

\begin{itemize}

	\item 4 events do not meet both sets of criteria and represent either
		electronics effects or atmospheric phenomena.

	\item 2 events have very special spatial or temporal structure, both cases are non-determinant due to our criteria.

	\item 4 events meet horizontal criteria.

	\item 3 events meet vertical criteria.

\end{itemize}

\begin{figure}[!ht]
  \centering
	\includegraphics[width=.7\textwidth]{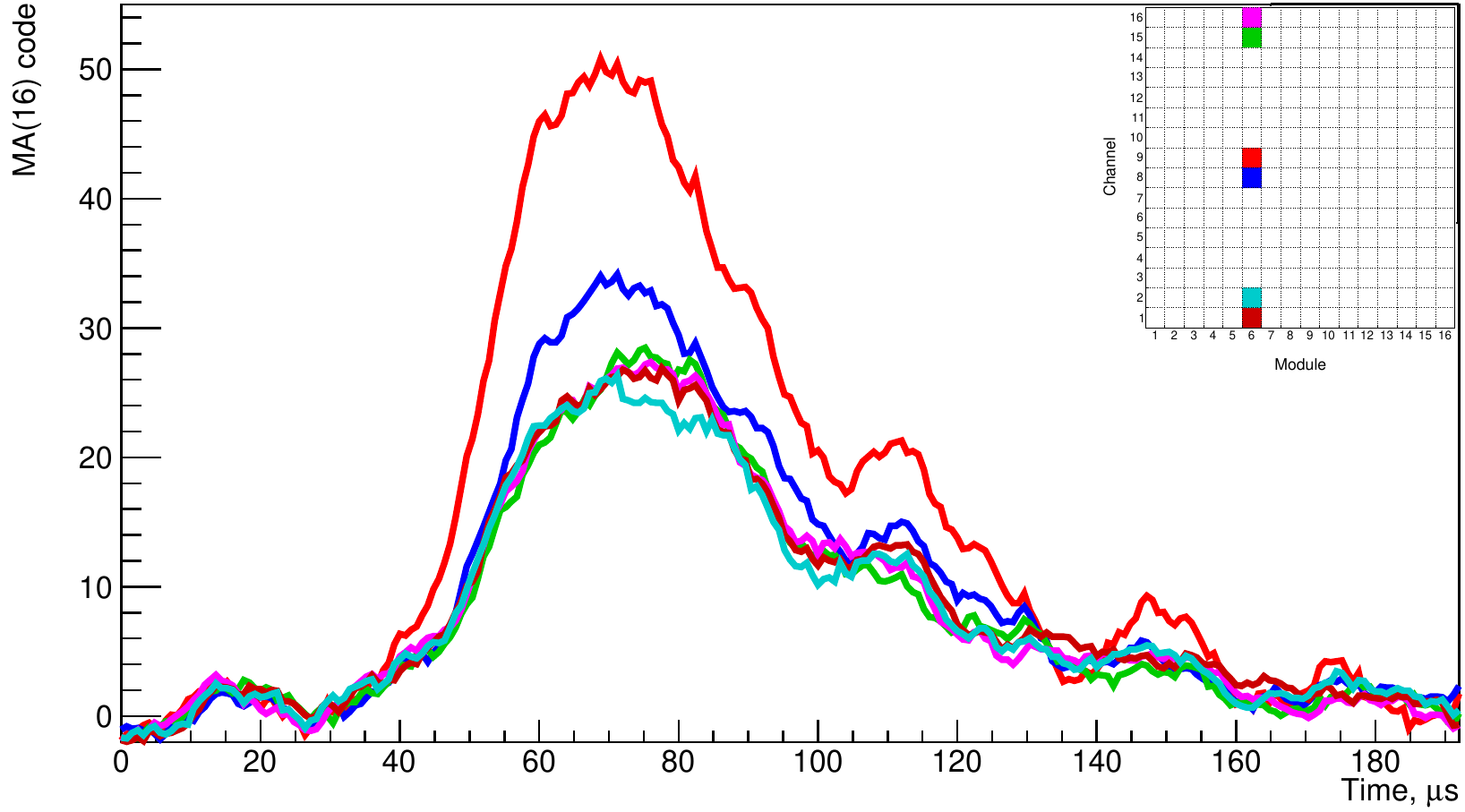}

	\caption{Moving average of ADC counts calculated for 16 time samples
	for the 6 brightest channels and their pixel map of event~\#1.}

	\label{Example1}
\end{figure}

An example of the first type of events is presented in
Fig.~\ref{Example1}.  Shown are values of the moving average (MA) of ADC
counts calculated for 16 time samples in the brightest pixels.  Here and
below, the MA curves in the figures are adjusted to have the same base
level of the signal.  For the event shown, all twelve active pixels are
located in the same PDM module and are hit simultaneously.  This cannot
be a vertical shower because of a large area of the illumination. For a
horizontal EAS, they should have a continuous signal producing track
which is not seen as well. This makes us conclude that this event is
more probably an effect of some cross-talk in the PDM.

\begin{figure}[!ht]
  \centering
	\includegraphics[width=.7\textwidth]{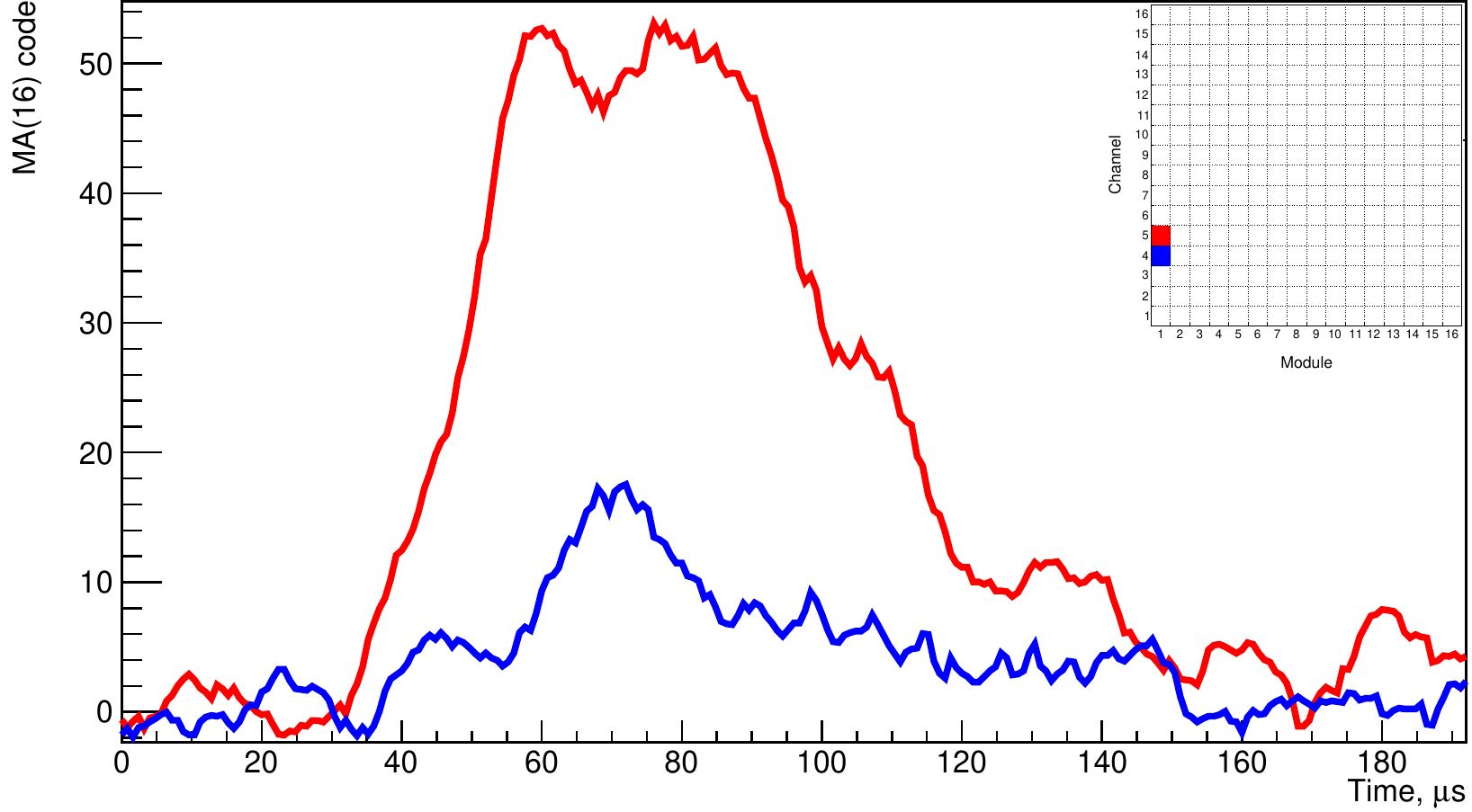}

	\caption{Moving average of ADC counts
	for active channels and their pixel map of event~\#2.}

	\label{Example2}
\end{figure}

An example of the second type of events
is provided by events~\#2 and~\#6, see Fig.~\ref{Example2},
but active pixels are located
at the very edge of the FOV, and it is difficult to make a definite
conclusion about a possible origin of the signal
since the PSF is largest in this area, so a source of the
signal can be outside the FOV.

\begin{figure}[!ht]
  \centering
	\includegraphics[width=.7\textwidth]{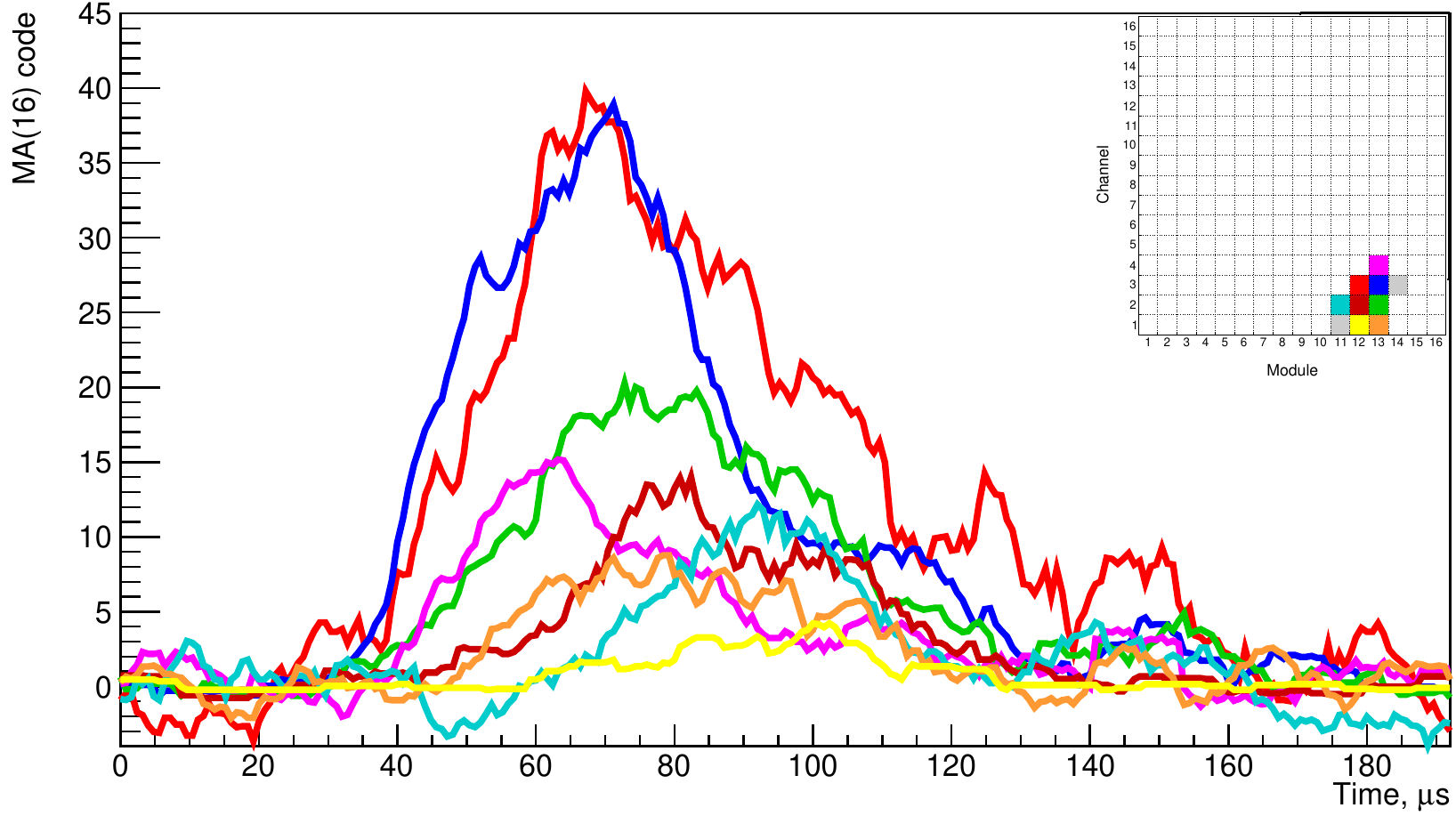}

	\caption{Moving average of ADC counts
	for active channels and their pixel map of event~\#3.}

	\label{Example3}
\end{figure}

One of the ``horizontal EAS'' candidates (event \#3) is presented in
Fig.~\ref{Example3}. It has the most interesting spatial-temporal
dynamics similar to what is expected for an EAS.  Active pixels are
grouped in an oblong spot, the shape of which might be a ``convolution''
of two factors--the mirror PSF and a linear track.

Main information about active pixels and temporal characteristics of the
signal are listed in Table~\ref{table3}. It can be seen from the
waveforms that characteristic duration of the signal is 70--100~$\mu$s
which is more that one can expect from a vertical EAS. The moment of
the maximum of the signal in each pixel has some shift from one pixel to
another.
This is an argument in favor of a horizontal (inclined) EAS
origin for this event.

\begin{table}[!ht]

\caption{Main parameters of event \#3: location of the active
	pixels (module and channel number), background level~$A_\text{bg}$
	and peak amplitude~$A_\text{max}$
	(in ADC codes), time~$T_\text{max}$
	of the peak from the beginning of the record,
	and the FDHM of the signal~$T_{1/2}$.
	All parameters are estimated basing on the moving average for 16 time
	samples.}

\begin{center}
\begin{tabular}{|c|c|c|c|c|}
	\hline
	md/ch&	$A_\text{bg}$ &	$A_\text{max}$ &	$T_\text{max},~\mu$s &
	$T_{1/2},~\mu$s\\
	\hline
13/4&	5.7&	15&	63&	36\\
12/3&	7.2&	40&	67.5&	47\\
13/3&	0.9&	38&	71.5& 41\\
13/2&	1.7&	20&	73.0&	49\\
13/1&	4.5&	8.6&	78&	42\\
12/2&	0.8&	14&	81&	41\\
11/2&	10.8&	 12.2&	92&	29\\
12/1&	0.2&	4.2&	100&	27\\
	\hline
	\end{tabular}
	\end{center}
	\label{table3}
\end{table}

It is important to mention that a possible thunderstorm activity was
studied in
the region of this event measurements. The Vaisala GLD360 ground
based lightning location network~\cite{Said2010,Said2013} did not
register any lightning strikes in a region with radius of 930~km and during
10~s period around the time of the TUS event. This provides a strong
support for a non-thunderstorm origin of the event.

Other EAS candidates of the horizontal type are events~\#7, \#9 and~\#10.

\begin{figure}[!ht]
  \centering
	\includegraphics[width=.7\textwidth]{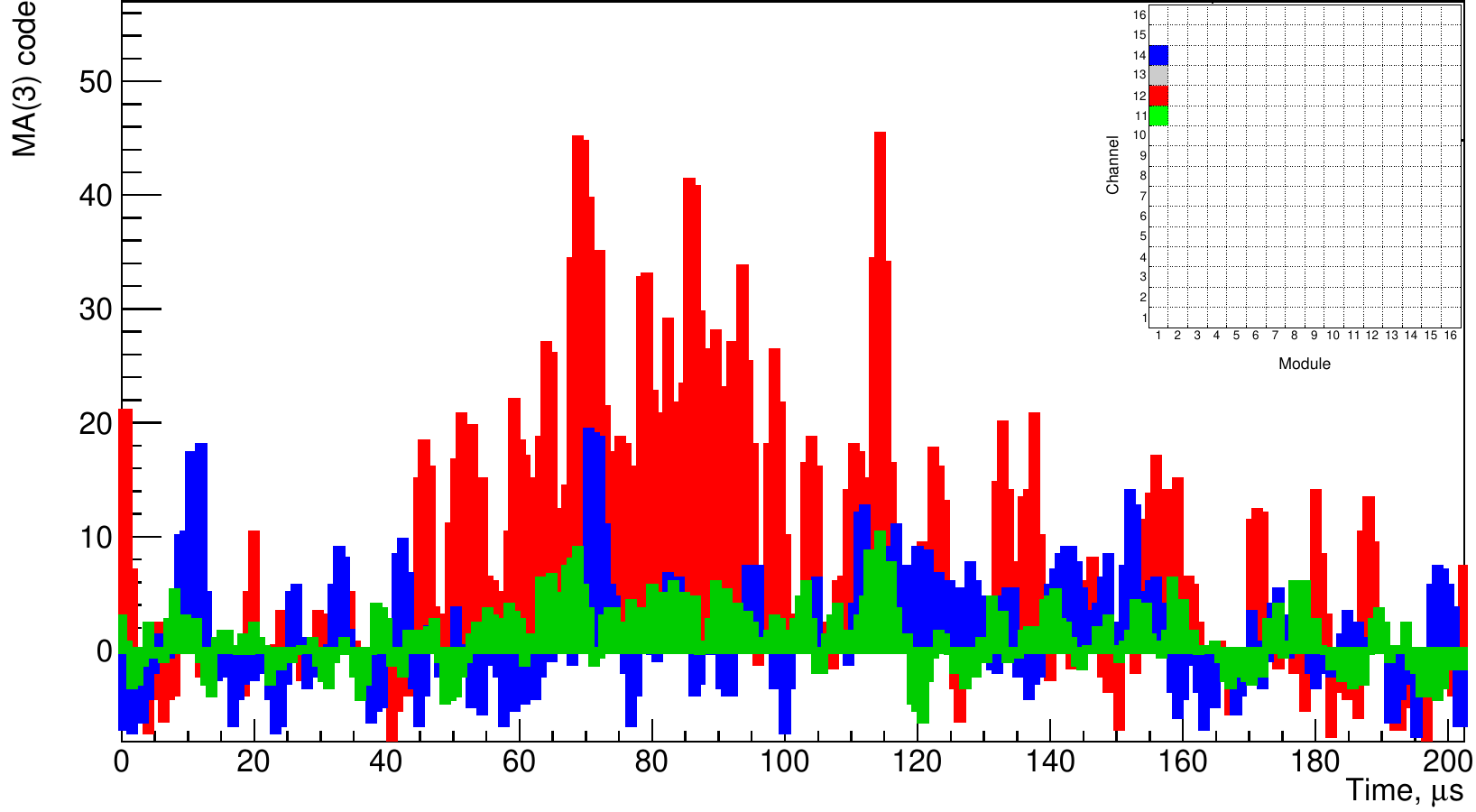}

	\caption{Moving average of ADC counts calculated for 3 time samples
	for active channels and their pixel map of event~\#4,
	a candidate for a vertical EAS.}

	\label{Example4}
\end{figure}

An example of an event that meets the criteria for a vertical EAS is
shown in Fig.~\ref{Example4}. Here in order to identify ``fine''
temporal structure we represent the moving average calculated for 3 time
samples, i.e., a moving average with time window $3\times0.8 = 2.4~\mu$s
only. The main part of the signal is located in one pixel, the other
active pixels are likely to be due to the PSF effects.  This signal has
$T_{1/2}\sim50~\mu$s, and it is very tempting to interpret the peak with
a time delay from the signal maximum equal to 25--30~$\mu$s as a
Cherenkov point. Other EAS candidates of the vertical type are
events~\#5 and~\#13.

\section{Discussion}

TUS, the first orbital detector of extreme energy cosmic rays, is
operating for approximately a year on board the Lomonosov satellite. In
a search for an EAS, generated by an EECR, it observes a large number of
background events of various origin that take place in the nocturnal
atmosphere of the Earth but they do not exclude a possibility to detect
a true EAS events at energies $\gtrsim70$~EeV.  A multi-level algorithm
for the search of EAS-like events was developed and applied to the TUS
data set.  As a result, 13 preliminary EAS candidates were selected and
analysed on the event by event basis according to the spatial-temporal
criteria for horizontal and vertical EASs.  Six of these candidates are
eliminated from the further analysis because they do not meet the
empirical criteria.  Other 7 events are considered as horizontal (4
events) or  vertical (3 events) EAS candidates.  The most interesting
for the further analyses are four candidates for horizontal EASs, which
demonstrate a movement of the image in the FOV. An analysis of their
temporal parameters is complicated by the large PSF and different
sensitivity of pixels, and needs more detailed and accurate simulations.
An influence of possible anthropogenic sources of illumination and
weather conditions must also be taken into account.  A more
detailed analysis of EAS candidate events found in the TUS detector data
is in progress, and its results will be reported elsewhere.

\section*{Acknowledgments}

The authors wish to thank Vaisala Inc.\ company for providing the data
on lightning strikes employed in the present study.  The TUS experiment
on board the Lomonosov satellite was realized within the Federal Space
Program of Russia with funding by the Russian Space Agency. The data
analysis is supported by RFBR grants No.\ 16-29-13065 and No.\
15-02-05498.  The Korean work
is supported by the National Research Foundation grants (No.\
2015R1A2A1A01006870 and No.\ 2015R1A2A1A15055344).


\end{document}